**Title:**

# Assessing fidelity-limiting factors and achieving single-qubit gate fidelity beyond 99.999% in driven silicon spin qubits

**Authors:**

Kenta Takeda[1, *], Akito Noiri[1], Takashi Nakajima[1], Leon C. Camenzind[1], Takashi Kobayashi[2], Giordano Scappucci[3], and Seigo Tarucha[1, 2 *]

**Affiliations:**

1 RIKEN Center for Emergent Matter Science (CEMS), RIKEN, Hirosawa 2-1, Wako-shi, 351-0198, Saitama, Japan.

2 RIKEN Center for Quantum Computing (RQC), RIKEN, Hirosawa 2-1, Wako-shi, 351-0198, Saitama, Japan.

3 QuTech and Kavli Institute of Nanoscience, Delft University of Technology, Lorentzweg 1, 2628 CJ Delft, Netherlands.

*Correspondence to: Kenta Takeda (kenta.takeda@riken.jp) or Seigo Tarucha (tarucha@riken.jp)

**Abstract:**

In semiconductor single-spin qubits, high-fidelity quantum gates have been demonstrated; however, achieving consistent performance remains challenging due to variations in driven qubit coherence, which is less explored than free-evolution coherence such as $T_2^*$. Here, we report single-qubit gate fidelities above 99.999%, achieved by dramatically extending the driven-spin coherence time and suppressing off-resonant driving effects that are detrimental to accurate fidelity benchmarking. We demonstrate that removing proximal reservoirs significantly enhances the spin-locking coherence time ($T_{1\rho}$), a critical metric for qubits under microwave driving. Furthermore, we reveal that in typical spin qubit setups using parity readout and rectangular pulses, off-resonant excitation of neighboring qubits causes substantial benchmarking artifacts. By optimizing device conditions to mitigate microwave-induced degradation and implementing spectrally tailored pulse shaping, we achieve a $\pi/2$ gate fidelity of 99.99920(2)%, with remaining errors primarily limited by incoherent noise. These results showcase the mechanisms that bound fidelity benchmarking in state-of-the-art silicon spin qubits and provide practical guidelines for achieving and verifying high fidelities in these systems.

**Main text**:

Semiconductor spin qubits offer a qubit platform compatible with modern semiconductor manufacturing[1–3], making them promising candidates for scaling toward fault-tolerant quantum computing with error-corrected qubits. Recent experiments in silicon qubits have reported high fidelities for fundamental qubit operations, such as single-qubit[4,5], two-qubit[6–8], and state preparation and measurement (SPAM)[9,10], with fidelities exceeding 99 %. Improving the fidelities beyond the bare minimum threshold required for error correction (approximately 99% for the surface code) is still essential to reduce the number of physical qubits needed to encode a logical qubit.

Among various silicon quantum-dot based spin qubit platforms (Si-MOS[2,5,10,11], Si/SiGe[4,7,12], etc.) with various encodings (spin-1/2[4,5,7], singlet-triplet[12], exchange-only[3], etc.), the spin-1/2 qubits implemented on $^{28}$Si/SiGe record single-qubit fidelity exceeding 99.99 %[13], along with two-qubit[6–8] and SPAM[14] fidelities exceeding 99 %. Single-qubit control in this architecture typically relies on electric dipole spin resonance (EDSR) utilizing an on-chip micromagnet. However, despite advances in material and manufacturing, as witnessed by common qubit metrics such as $T_2^*$, single-qubit fidelities reported in many experiments remain at around 99.9%[1,4,7,9]. While our recent work on a five-qubit array demonstrated single-qubit fidelities above 99.99%[13], that study focused primarily on the simultaneous operation of multiple spin qubits, leaving the physical mechanisms behind the enhanced fidelity largely unexplored. Identifying the limiting factors for both gate fidelity and its accurate measurement is therefore crucial for reproducibly achieving high fidelities in future experiments.

Here, we demonstrate single-qubit π/2 rotation fidelities as high as 99.999 % through a detailed characterization of the limiting factors in randomized benchmarking (RB) experiments. We identify additional decoherence caused by proximal reservoirs, as well as microwave-induced frequency shifts, as detrimental mechanisms limiting the RB fidelity. Furthermore, we find that, when rectangular pulses and parity readout are combined, off-resonant driving of the idle qubit which participates in the parity readout can result in an enhanced decay of the return probability in single-qubit RB experiments. By employing pulse shaping to enable reliable RB measurements, we characterize the dependence of the gate fidelity on the reservoir proximity, operating temperature, and microwave amplitude. Under optimized conditions including the elimination of a proximal reservoir, operation at an elevated temperature of 100 mK[15], and the use of spectrally narrow microwave pulses[16], we achieve a π/2 rotation fidelity of 99.99920(2) %, predominantly limited by decoherence.

**Experimental setup and basic device properties**

Figure 1a shows a scanning electron microscope image of the device, in which we can host up to five coupled quantum dots. In this study, we focus on a double quantum dot under gates P1 and P2. An additional quantum dot formed under gate SP1 is used as a charge sensor, and its conductance is measured by radio-frequency reflectometry. A cobalt micromagnet is fabricated on top of the gate stack, enabling addressable single-qubit control with EDSR[17] and two-qubit conditional gate[18]. The double quantum dot is operated in the (3,1) charge state, where numbers ($N_1$, $N_2$) denote the charge occupation of quantum dots formed under P1 and P2, respectively. In what follows, we refer to the spin qubits formed under gate P1 (P2) as Q1 (Q2).

Single-shot spin readout is performed via spin-to-charge conversion using the Pauli spin blockade (PSB) mechanism[19]. In the PSB regime, the antiparallel spin states ($|\uparrow\downarrow\rangle$, $|\downarrow\uparrow\rangle$) are transferred to the (4,0) charge configuration, whereas the triplet spin states ($|\downarrow\downarrow\rangle$ and $|\uparrow\uparrow\rangle$) remain in the (3,1) configuration. By discriminating between the (4,0) and (3,1) charge states, we implement a parity (ZZ) spin readout. Our experimental sequence begins with a PSB measurement (M1 in Fig. 1b), which we use to post-select the S(4,0) initial state. Then, we apply a voltage ramp to the (3,1) charge configuration to initialize the $|\uparrow\downarrow\rangle$ state. This voltage ramp is carefully calibrated to avoid unwanted population transfer to the low-lying valley-excited or other spin states. Spin manipulation is performed at the charge-symmetry point to minimize residual exchange coupling, which we estimate to be well below 10 kHz (see Supplementary Fig. 1). Exchange coupling is controlled by pulsing the virtual barrier gate vB1 while maintaining the charge-symmetry condition. After the spin manipulation, a reverse voltage ramp returns the system to the PSB configuration for readout. We then obtain even-parity probability $P_{\mathrm{even}}$ from the second PSB readout (M2 in Fig. 1b).

Figure 1c shows a typical spin resonance spectrum measured at an external magnetic field $B_{\mathrm{ext}} =$ 0.48 T. Because the microwave is applied via a common driving gate (SG1), and due to the nature of the readout scheme, resonance peaks from both spins appear in the spectrum. The two peaks are separated by $\Delta f$ =126.6 MHz due to the magnetic field gradient induced by the micromagnet. Figure 1d summarizes the basic qubit characteristics. The $T_2^*$, Hahn echo decay time ($T_2^{\mathrm{H}}$), and Carr-Purcell-Meiboom-Gill (CPMG) echo time ($T_2^{\mathrm{CPMG}}$) represent coherence times under free evolution, while Rabi decay time ($T_2^{\mathrm{Rabi}}$), spin-locking decay time ($T_{1\rho}$) characterize coherence under continuous microwave driving (see Methods and Supplementary Fig. 2 for details). These metrics contribute to the RB decay, depending on how the randomized gates are applied. For example, a modest RB fidelity of 99.9% result in an 1/e decay length of 1000 Clifford gates. Using the average Clifford gate time of 325 ns, this corresponds to an effective RB 1/e decay time ($T_2^{\mathrm{RB}}$) of approximately 325 μs. This timescale is much longer than the typical free-evolution coherence times ($T_2^* < 10$ μs), suggesting substantial decoupling effects from microwave driving.

**Influence of proximal reservoir**

Fig. 1d shows comparable performance metrics for Q1 and Q2, with the striking exception of their $T_{1\rho}$ values. The primary difference between these two qubits is their spatial proximity to the charge sensor and its reservoirs, although the closer qubit Q1 does not directly tunnel couple to any reservoirs. To assess the potential influence of nearby reservoirs on coherence times, especially $T_{1\rho}$, we perform a control experiment in which the right side of the device is configured as an extended reservoir. By tuning the voltages on gates P3-P5 and B3-B5, we control the proximity of this reservoir to Q2 (Fig. 2a-c). To exclude direct electron tunneling events between Q2 and this extended reservoir, the tunnel rate is kept below 0.2 Hz, which is much slower than the timescale of spin dynamics. Figure 2d shows the measured $T_{1\rho}$ values for Q2 for the different reservoir configurations. As the reservoir is brought closer to the qubit, $T_{1\rho}$ degrades significantly via the relation $T_{1\rho}^{-1} = \frac{1}{2}T_1^{-1} + \frac{(2\pi)^2}{2}S_{\parallel}(f_{\mathrm{Rabi}})$, where $T_1^{-1} = \frac{(2\pi)^2}{2}S_{\perp}(f_{\mathrm{Larmor}})$ is the longitudinal relaxation (depolarization) rate and $f_{\mathrm{Larmor}}$ is the Larmor frequency[20]. Here, $S_{\parallel}(f)$ and $S_{\perp}(f)$ are the power spectral densities of the longitudinal and transverse noise components at frequency of $f$, defined in units of $\mathrm{Hz}^2/\mathrm{Hz}$. The reduction in $T_{1\rho}$ suggests that either $S_{\perp}(f_{\mathrm{Larmor}})$, $S_{\parallel}(f_{\mathrm{Rabi}})$, or both significantly increase when the reservoir is brought closer to the qubit. This proximity dependence is likely explained by the direct qubit-reservoir capacitive coupling, as discussed in the Supplementary Information.

Figure 2e summarizes additional qubit metrics for the same reservoir configurations. In contrast to $T_{1\rho}$, dephasing times such as $T_2^*$ and $T_2^{\mathrm{CPMG}}$ show no systematic reduction as the reservoir is brought closer to the qubit, indicating that the reservoir does not induce a substantial increase in low-frequency noise. This behavior suggests that an increase in charge noise caused by the activation of charge fluctuators is not the dominant mechanism. Instead, weakly frequency-dependent voltage noise from the reservoir, such as Johnson-Nyquist noise, is consistent with the measurement results. Such noise is expected to be finite even in the absence of microwave driving, but it also depends on the electron temperature of the reservoir and can therefore be enhanced by drive-induced heating.

Further insight can be obtained by measuring $T_1$ under a microwave driving, where we find that it is significantly longer than $T_{1\rho}$ even in the presence of the nearby reservoir (Fig. 2e). The large discrepancy between $T_1$ and $T_{1\rho}$ indicates that $S_{\parallel}(f_{\mathrm{Rabi}})$ increases much more rapidly than $S_{\perp}(f_{\mathrm{Larmor}})$ as the reservoir is brought closer. This behavior is consistent with the quantum-dot/reservoir geometry and the anisotropy of the magnetic-field gradient. Voltage fluctuations from the reservoir predominantly displace the quantum dot along the array direction (x-axis in Fig. 1a). Our micromagnet simulation (Supplementary Fig. 3) suggests that the relevant field gradients for $T_1$ ($\mathrm{d}B_{\mathrm{x}}/\mathrm{d}x$ and $\mathrm{d}B_{\mathrm{y}}/\mathrm{d}x$) are negligible compared with the one relevant for $T_{1\rho}$ ($\mathrm{d}B_{\mathrm{z}}/\mathrm{d}x$). This anisotropy accounts for the selective enhancement of longitudinal noise that shortens $T_{1\rho}$, while explaining why $T_1$ remains long despite the nearby reservoir.

**Microwave-induced frequency shift**

In addition to the enhanced decoherence, microwave driving can also cause heating-induced shifts in the qubit frequency[15,21]. This effect detunes the qubit frequency from its free-evolution value, resulting in coherent errors. Figures 3a, b show the measured resonance peak positions of Q2 as a function of the duration of an off-resonant pre-heat microwave pulse at a base temperature of 30 mK. We observe a clear shift of the resonance frequency; however, this shift is significantly smaller than values reported in the literature (negligibly small frequency shift for $20\pi$ pulses compared to $0.5-1$ MHz shift reported in Ref.[15]). Nevertheless, such a frequency shift is detrimental to quantum protocols requiring a substantial number of pulses, for example, RB for high-fidelity gates and quantum error correction cycles. We also note that no temperature sweet spot[15] is observed for the frequency shift.

One approach to mitigating this effect is to raise the mixing chamber temperature ($T_{\mathrm{MXC}}$), which increases the available cooling power and thereby reduces the instantaneous change in device temperature during microwave irradiation[15]. For example, operation at 100 mK suppresses the frequency shift to below 50 kHz for a 2-ms pulse, which corresponds to 24,000 $\pi/2$ pulses for $f_{\mathrm{Rabi}}=3$ MHz (Fig. 3c,d). Figure 3e shows the $T_{\mathrm{MXC}}$ dependence of the frequency shift from the base temperature value. In addition, we find that the absolute value of the frequency shift upon increasing $T_{\mathrm{MXC}}$ is comparable to that reported in Ref.[15], even though the pulse-induced temporal frequency shift is significantly smaller. This may imply higher cooling power in our device, leading to a smaller transient increase in device temperature than in Ref.[15], potentially arising from differences in materials or the experimental configuration such as thermal anchoring.

**Parity spin readout and pulse shaping**

Parity spin readout in quantum-dot spin qubits imposes stringent frequency selectivity on qubit control. Although rectangular pulses (Fig. 4a) are commonly used for their simplicity and efficient power delivery, their high-amplitude spectral sidelobes (Fig. 4b) lead to off-resonant driving of nearby qubits. While this effect is usually not a major concern as long as considering the single-qubit subspace, it becomes problematic in randomized benchmarking (RB) or multi-qubit characterizations. In RB, small random off-resonant rotations can accumulate over long gate sequences, resulting in significant population transfer in the off-resonant qubit. Because parity readout outcomes depend on the joint spin configuration, this additional depolarization enhances the observed RB decay, which no longer reflects only the qubit fidelity of interest.

To isolate this effect, we directly compare RB measurements of Q2 using two different readout schemes: unconditional IZ readout and ZZ parity readout. The IZ readout is implemented by combining a controlled rotation with the parity readout (Fig. 4c; see Supplementary Information for details). Denoting the depolarizing parameter for Q2 as $p$ and the off-resonant driving depolarizing parameter for Q1 as $q$, the ZZ readout yields a combined decay characterized with $p + q$, whereas the IZ readout yields a decay characterized by $p$. As shown in Fig. 4d, the RB data obtained with the ZZ readout (lowest panel) decays noticeably faster than that for IZ readout. If one were to just characterize the ZZ component, it might be misled to reduced gate fidelity. In reality, the measurement is distorted by the additional depolarization of Q1 due to off-resonant driving. These results demonstrate that the high gate fidelity cannot be accurately measured with the combination of rectangular pulse and ZZ readout.

To further confirm the influence of pulse spectral broadening, we perform RB for various gate times $t_{\mathrm{g}}$. The off-resonant driving is maximized (minimized) for rotation angles that are odd (even) multiples of $\pi$, known as the synchronization condition. In Fig. 4e, we measure the RB decay per Clifford gate for various $t_{\mathrm{g}}$ sampled based on the synchronization condition (see Supplementary Information for details). We observe the expected periodicity of $(\Delta f)^{-1} = 7.9$ ns for the ZZ readout, which nearly vanishes with the IZ readout, thereby confirming that off-resonant driving is indeed the limiting factor in the enhanced RB decays. Numerical simulations (blue band in Fig. 4e, see Methods for details) show good agreement with the experimental data. Furthermore, simulations for Gaussian pulses predict negligibly small values of $q$ over the range of $t_{\mathrm{g}}$ used in subsequent RB experiments (Supplementary Fig. 5). Consequently, we employ Gaussian pulses and ZZ readout for the single-qubit RB experiments that follow.

**RB characterizations and error budgeting**

The characterizations above identify isolation from the reservoir and operating temperature as key parameters for achieving high-fidelity single-qubit control in this system. Additionally, shaped control pulses enable reliable extraction of the single-qubit depolarizing parameter, even with parity spin readout. Figure 5a shows several representative RB measurements obtained under different operating conditions, demonstrating significant variations in RB fidelity. Consistent with the coherence time measurements, isolating the qubit from the reservoir leads to significant improvement (Fig. 5b). Fig. 5c shows the temperature dependence of the gate infidelity, which exhibits a minimum at around 100 mK. At higher temperatures, the fidelity decreases, likely due to increased charge noise and reduced relaxation times at elevated temperatures. The driving strength dependence measured at 100 mK shows that the gate fidelities consistently exceed 99.999 % around a peak $f_{\mathrm{Rabi}} = 3$ MHz (Fig. 5d; note that peak $f_{\mathrm{Rabi}}$ is 1.87 times larger than the average $f_{\mathrm{Rabi}}$).

To infer the residual performance limiting mechanisms, we analyze the RB experiment using purity measurement (Fig. 5e). Unlike standard RB, which measures only the return probability at the end of sequences, one can measure the density matrix $\rho$ by using additional projection pulses before the readout. The decay of the purity, $\mathcal{P} = \mathrm{Tr}(\rho^2)$, reflects coherence loss, and comparison with the RB decay separates coherent and incoherent contributions to the total RB error[22,23]. Notably, the experimental and post-processing costs of this approach are comparable to those of standard RB, making it a practical method for assessing fidelities of 99.99 % and beyond. Figure 5f shows the comparison of RB decay and purity decay for Q2. From these measurements, we obtain the RB decay rate (per $\pi/2$ gate) of $\varepsilon_{\mathrm{RB}} = 8.6(3) \times 10^{-6}$ and the purity decay rate of $\varepsilon_{\mathrm{PB}} = 7.1(2) \times 10^{-6}$. From these, we obtain a unitary error per $\pi/2$ rotation is as low as $1.5(4) \times 10^{-6}$. While a measurable residual coherent error exists, the result implies that the error is predominantly caused by decoherence.

**Conclusion and outlook**

In summary, we have characterized a $^{28}$Si/SiGe spin-qubit device under various operating conditions, including the reservoir proximity, operating temperature, and control pulse shape. By optimizing these factors, we demonstrate $\pi/2$ gate fidelities exceeding 99.999 %, limited by decoherence. While this level of fidelity is well-positioned for integration into larger-scale processors, our results indicate that microwave-induced heating effects are significant even for operating a single spin. Although mitigation schemes exist[13], scaling to larger systems will require substantial reduction of heat dissipation, for example via baseband operation[24] or low-frequency EDSR[25]. Furthermore, while reservoirs are essential for tuning up quantum-dot charge states, reduced proximity to reservoirs could be efficiently achieved via a sparse, shuttling-based architecture[26,27]. Taken together, our results provide concrete design guidelines for realizing large-scale silicon spin-qubit systems with optimal performance.

**Methods**

General

The $^{28}$Si/SiGe quantum-dot device is nominally identical to the one used in Ref.[13], except that a different singulated die was used. The electrons are accumulated in a buried $^{28}$Si quantum well with 800 ppm residual $^{29}$Si using three-layer overlapping aluminum gates. The aluminum gates, as well as the cobalt micromagnet, are fabricated using standard electron-beam lithography and lift-off techniques (see Ref.[13] for details). All measurements are performed in a dilution refrigerator (Bluefors XLD1000sl) with a base temperature of 20 to 30 mK. The sample is glued with GE Varnish to the ground plane of a custom printed circuit board (PCB), which is thermalized to the mixing chamber. From the thermal broadening of the dot-reservoir charge transition, we estimate the base electron temperature to be approximately 50 mK (see Supplementary Figure 6). At $T_{\mathrm{MXC}}=100$ mK, we measure an electron temperature of approximately 100 mK.

Control electronics

The DC voltage is supplied by a multichannel DAC (Qblox D5a) and combined with the high-frequency signal via resistive bias-tees (R=510 kΩ and C=100 nF) on the sample PCB. The plunger and barrier gates, including those for the charge sensors, are connected to a baseband arbitrary waveform generator (AWG, Keysight M5301A). A digital pre-distortion filter is implemented on its internal FPGA to correct the pulse distortion caused by the group delay of lowpass filter. Gate SG1 is connected to a microwave AWG (Keysight M5300A), which directly synthesizes the microwave signal without analog IQ mixing. All AWG modules are installed in the same PXI chassis (Keysight M9046A) and are synchronized to its internal clock. For the baseband pulses, a multichannel, custom room-temperature variable-gain voltage combiner is used to compensate for pulse distortion caused by the cryogenic bias-tee[28].

Readout

For radio-frequency reflectometry, one of the ohmic contacts (upper left in Fig. 1a) for the charge sensor quantum dot is coupled to an LC tank circuit, which consists of a surface mount inductor (Coilcraft 1206CS-122X) and a parasitic capacitance. The tank circuit is operated at its resonance frequency of 178.4 MHz. The carrier signal is generated by an AWG (Keysight M5301A) and applied to the ohmic contact of the charge sensor via attenuators and a directional coupler. This carrier signal is applied only during the readout stages (M1 and M2 in Fig. 1b). Except for these readout stages, the charge sensor is pulsed to the Coulomb blockade regime. The reflected signal is first amplified by a cryogenic amplifier (Cosmic Microwave Technology CMT-BA1) mounted at the 4 K stage of the dilution refrigerator. At room temperature, the signal is further amplified and then digitized (Keysight M5200A). The internal FPGA in the digitizer is used for demodulating the signal to the baseband in-phase voltage $V_{\mathrm{rf}}$. We integrate $V_{\mathrm{rf}}$ for 10 μs to obtain a single-shot data, resulting in a signal-to-noise- ratio (SNR) of ~14. While this SNR enables near unity state discrimination fidelity, the state mapping error for PSB practically limits typical RB visibilities (or SPAM fidelity) to ~ 95 %. We observe that operation at temperatures higher than 200 mK degrades the readout fidelities due to reduced SNR and enhanced relaxation. We typically repeat 200 to 400 shots to obtain a single probability. Post-selecting the S(4,0) initial state roughly halves the number of usable shots for calculating $P_{\mathrm{even}}$.

Coherence time metrics

Supplementary Fig. 2 shows the pulse sequences used for characterizing the values reported in Fig. 1d. The $T_1$ values are not reported; however, we do not observe a significant relaxation on a 10 ms timescale. The $T_2^*$ is determined by integrated low-frequency noise and, in this device, we believe it is dominated by nuclear spins due to the distinct temperature dependence from charge-noise dominated metrics (see Supplementary Fig. 7). The spin echo sequences used to measure $T_2^{\mathrm{H}}$ and $T_2^{\mathrm{CPMG}}$ decouple such nuclei-induced low-frequency noise and are likely limited by charge noise. Since microwave driving is nearly continuously applied during the RB experiments, these free-evolution dephasing times are not sufficient to characterize the RB decay. The $T_{1\rho}$ time, known as spin-locking time, is linked to high-frequency noise at $f_{\mathrm{Rabi}}$ under microwave driving. There are some practices utilizing $T_{1\rho}$ measurement for noise spectroscopy[29], however, its utility here is limited because the noise environment strongly depends on the heating (or $f_{\mathrm{Rabi}}$). Experimentally, we observe monotonic reduction of $T_{1\rho}$ as $f_{\mathrm{Rabi}}$ is increased (Supplementary Fig. 8). Meanwhile, $T_2^{\mathrm{Rabi}}$ involves contributions from both low-frequency noise and noise at $f_{\mathrm{Rabi}}$ (both driving amplitude and qubit energy fluctuations). Since RB sequences result in both spin-locking and Rabi-like time evolutions, these metrics contribute to the RB decay in a nontrivial manner.

Randomized benchmarking

The single-qubit Clifford gates are decomposed into $\pi/2$ rotations ($\mathrm{X}_{\pi/2}$ and $\mathrm{Y}_{\pi/2}$) as in Ref. [13] (the identity gate is implemented as $\mathrm{X}_{\pi/2}\mathrm{X}_{\pi/2}\mathrm{X}_{\pi/2}\mathrm{X}_{\pi/2}$). This decomposition results in an average of 3.25 $\pi/2$ rotations per Clifford gate. 0 to 10 ns waiting time is inserted between the $\pi/2$ pulses. The microwave amplitude and frequency are calibrated at intervals of roughly 5 min. Such automated calibration was not implemented for the gate voltages, since those are typically stable over a time scale of a few days. After the final microwave pulse in each sequence, a 0.5 ms delay is added before ramping gate voltages to the PSB regime for readout. We measure the return probabilities $p_\uparrow(m)$ and $p_\downarrow(m)$ using two types of recovery Clifford gates, targeting $|\uparrow\rangle$ and $|\downarrow\rangle$ final states, respectively. The sequence fidelities shown in Figs. 4, 5 are obtained as $F(m) = p_\uparrow(m) - p_\downarrow(m)$, which decay to zero for a completely randomized result. For each RB experiment, we average over 32 random Clifford gate sequences.

Error budgeting (purity benchmarking)

While gate set tomography provides detailed information about error channels, evaluating errors as small as $10^{-4}$ or below results in prohibitively long experimental runtime[13]. We adopt purity benchmarking (PB) as a low-cost alternative to assess the residual fidelity-limiting mechanism. Note that both GST and PB, if performed in the single-qubit space, fail to account for off-resonant rotations of the reference qubit. For implementing PB, in addition to the standard projective measurement to obtain $\langle\sigma_\mathrm{z}\rangle$, we apply pre-rotations to measure the Pauli expectation values $\langle\sigma_\mathrm{x}\rangle$ and $\langle\sigma_\mathrm{y}\rangle$. The purity parameter is then computed as $\mathcal{P} = \mathrm{Tr}(\rho^2) = \langle\sigma_\mathrm{x}\rangle^2 + \langle\sigma_\mathrm{y}\rangle^2 + \langle\sigma_\mathrm{z}\rangle^2$ and the decay of $\sqrt{\mathcal{P}}$ is compared with the RB decay. For depolarizing errors, the decay of $\sqrt{\mathcal{P}}$ is expected to be equivalent to RB decay. Since coherent errors caused by unitary evolution do not affect purity, generally purity error is smaller than RB error ($\varepsilon_\mathrm{PB} < \varepsilon_\mathrm{RB}$). From these measurements, the coherent error can be quantified as $\varepsilon_\mathrm{RB} - \varepsilon_\mathrm{PB}$. The two traces in Fig. 5f are obtained from the same run of experiment so that the slow environmental drift does not affect the result, with averaging performed over 16 random Clifford gate sequences.

Statistical analysis

Unless otherwise noted, the error bar represents $1\sigma$ from the mean, obtained by least-square fit of the data.

**Data availability**

The data that support this work will be made available at Zenodo repository at {URL}.

**Code availability**

The codes used for data processing and numerical simulation are available from the corresponding authors upon reasonable request.

**Acknowledgements**

We thank Reiko Kuroda for assistance with sample fabrication and Soichiro Teraoka for assistance with measurement hardware preparation. Fabrication of the samples used in this work was performed at the Nanoscience Joint Laboratory, RIKEN. This work was supported financially by JST Moonshot R&D grant number JPMJMS256H; JSPS KAKENHI grant numbers 23H05455 (K.T.), 23K26483 (A.N.), 26K00642 (L.C.C.), and 22H01160 (T.K.); JST PRESTO grant number JPMJPR23F8 (A.N.); RIKEN Incentive Research Project 202601080071 (L.C.C.).

**Author Contributions**

A.N. fabricated the device. K.T. performed the measurements. A.N., T.N., L.C.C., and T.K. contributed data acquisition and discussed the results. G.S. developed and supplied the $^{28}$Si/SiGe heterostructure. K.T. wrote the manuscript with inputs from all co-authors. S.T. supervised the project.

**Competing interests**

The Authors declare no Competing Financial or Non-Financial Interests.

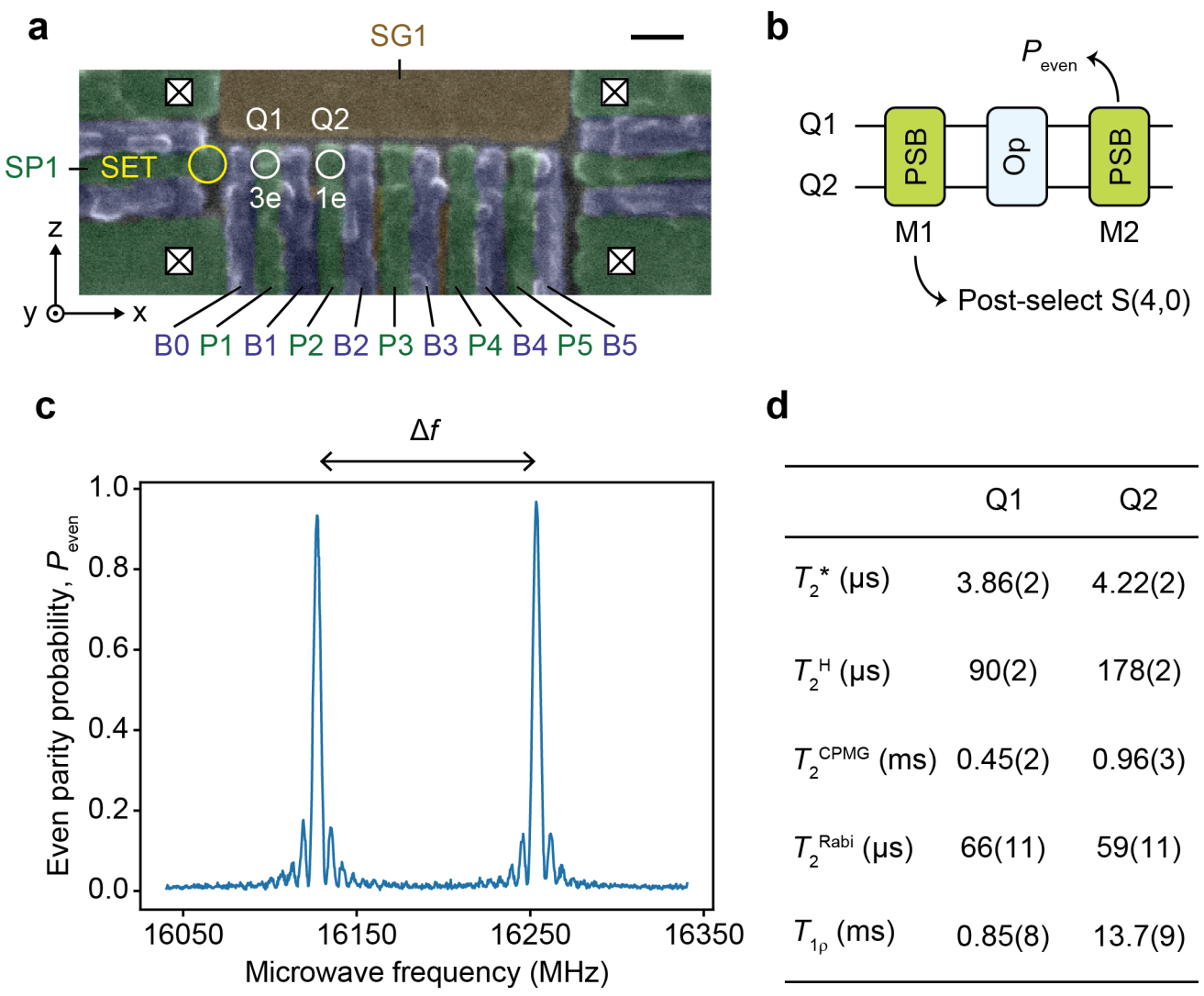

| | Q1 | Q2 |
|---|---|---|
| $T_2^*$ (μs) | 3.86(2) | 4.22(2) |
| $T_2^{H}$ (μs) | 90(2) | 178(2) |
| $T_2^{CPMG}$ (ms) | 0.45(2) | 0.96(3) |
| $T_2^{Rabi}$ (μs) | 66(11) | 59(11) |
| $T_{1\rho}$ (ms) | 0.85(8) | 13.7(9) |

Fig. 1, Basic device properties. **a**, False-color scanning electron microscope image of the device. The khaki, green, and blue regions represent the screening, plunger, and barrier gate layers, respectively. Scale bar, 100 nm. The boxes indicate ohmic contacts. The upper-left contact is connected to the tank circuit for reflectometry, while the others are directly grounded to the ground plane of the sample holder. The external magnetic field is applied along the z-axis. **b**, Schematic of measurement sequence. M1 and M2 denote PSB measurements. Op represents the operation stage, during which microwave and baseband pulses are applied. Although redundant except for the first cycle of the pulse sequence, M1 and M2 are performed in each cycle. **c**, EDSR spectrum of the two spin qubits measured using PSB parity spin readout. The microwave signal has a rectangular envelope with approximately 3 MHz Rabi frequency, although the microwave amplitude reaching the device is frequency dependent due to transmission properties. **d**, Qubit coherence times. The errors represent $1\sigma$ from the mean. The $T_2^*$ data are integrated for $\approx$1600 seconds. For the CPMG measurement, we apply 32 refocusing pulses. The $T_2^{\mathrm{Rabi}}$ and their errors are obtained by averaging 16 independent runs.

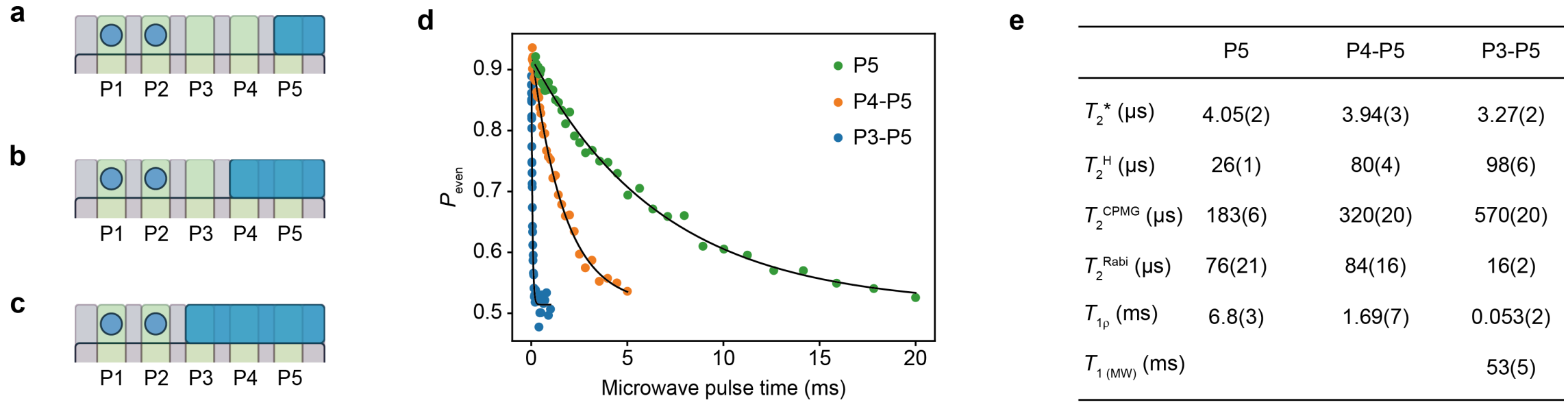

| | P5 | P4-P5 | P3-P5 |
|---|---|---|---|
| $T_2^*$ (μs) | 4.05(2) | 3.94(3) | 3.27(2) |
| $T_2^H$ (μs) | 26(1) | 80(4) | 98(6) |
| $T_2^{CPMG}$ (μs) | 183(6) | 320(20) | 570(20) |
| $T_2^{Rabi}$ (μs) | 76(21) | 84(16) | 16(2) |
| $T_{1\rho}$ (ms) | 6.8(3) | 1.69(7) | 0.053(2) |
| $T_{1\,(MW)}$ (ms) | | | 53(5) |

Fig. 2, Qubit characterization with an adjacent reservoir. The measurements are performed for Q2 (see Supplementary Fig. 4 for Q1 characterization results). **a-c**, schematics of reservoir configurations. Blue areas indicate regions where a two-dimensional electron gas is accumulated. The two-dimensional gas is grounded via the ohmic contact for the right charge sensor. The voltages are all set to 0.7 V, which is well above the accumulation threshold. The voltage on the barrier gate between P2 and P3 (B3 in Fig. 1a) is kept constant at 0.2 V, resulting in a 0.2 Hz tunnel rate between Q2 and the reservoir when the reservoir extends below P3. **d**, $T_{1\rho}$ measured for different reservoir configurations, showing drastic dependence on the reservoir configurations. **e**, Summary of coherence metrics for different reservoir configurations. $T_{1(\mathrm{MW})}$ refers to the relaxation time under an off-resonant microwave pulse (+100 MHz offset from the Q2 resonance frequency) with an amplitude corresponding to $f_{\mathrm{Rabi}}$ = 3 MHz.

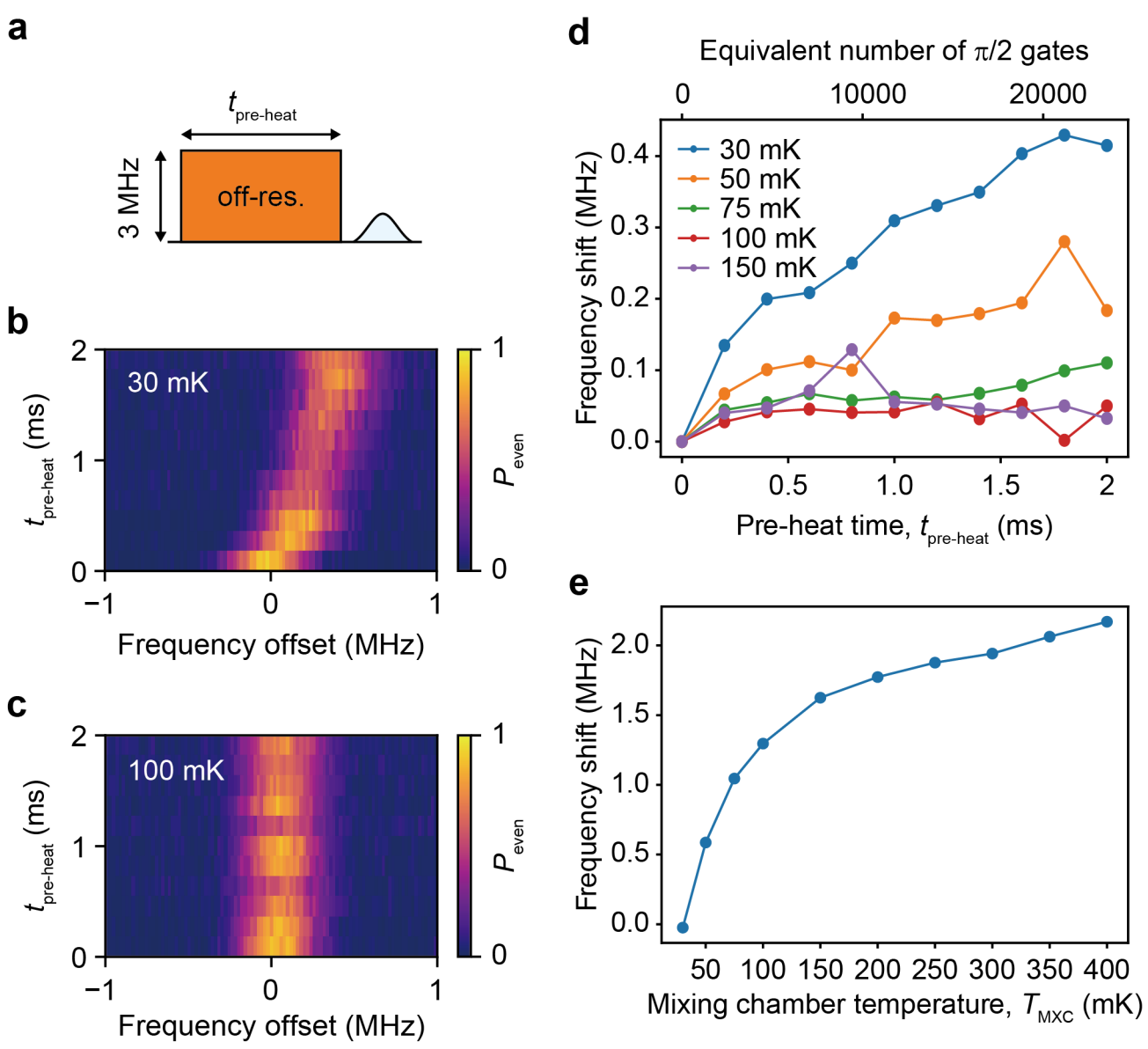

Fig. 3, Heating effect and temperature dependence. **a**, Schematic of heating effect measurement. The off-resonant pulse is detuned by +100 MHz from the resonance frequency. The 3 MHz drive amplitude is inferred from Rabi oscillation for resonant pulse, thus may not be representing the exact drive amplitude for off-resonant pulse. **b**, Frequency shift measured at $T_{\text{MXC}} = 30$ mK. **c**, Frequency shift measured at $T_{\text{MXC}} = 100$ mK. **d**, Frequency shift as a function of pre-heat time, measured at various mixing chamber temperatures. The frequencies are obtained by fitting the data with Gaussian function. The fitting error ($1\sigma$) is smaller than the size of markers, while the scattering of frequency comes from slow drift of environment. **e**, Frequency shift from the resonance frequency at the base temperature. Note that these frequency shifts are stationary values obtained from standard Ramsey-based measurements.

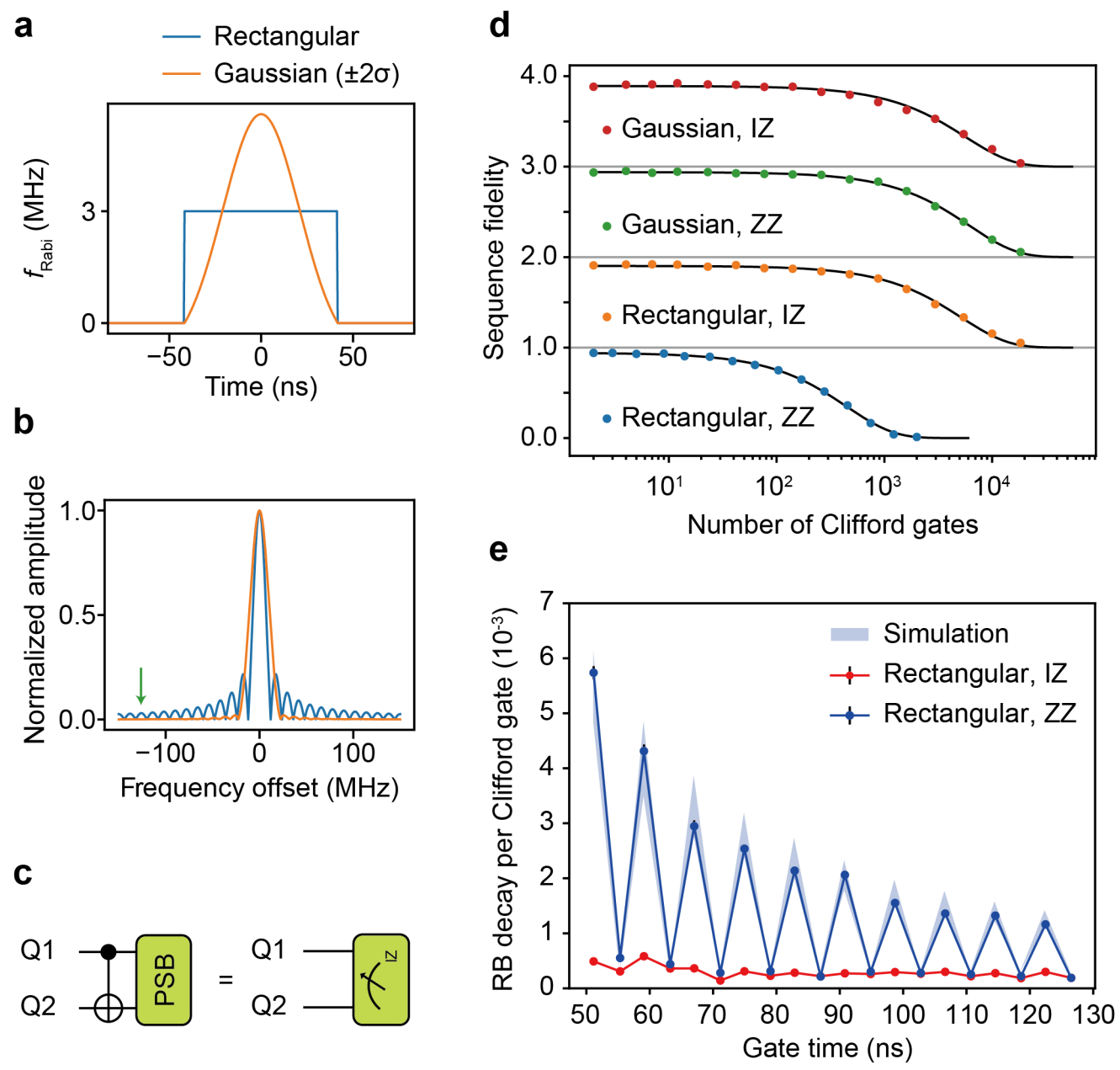

Fig. 4, Pulse shape comparison and characterization of off-resonant driving. Measurements are performed with the right reservoir depleted and at $T_{\mathrm{MXC}} = 30$ mK. **a**, Time-domain comparison of rectangular pulse and Gaussian pulse envelopes. Both pulses are designed to implement a $\pi/2$ gate with a duration of $0.25 \times (3\ \mathrm{MHz})^{-1} \sim 83.3$ ns. The Gaussian pulse is truncated at $2\sigma$, where $\sigma$ is the standard deviation. To mitigate pulse sidelobes caused by simple truncation, the entire pulse is offset so that it starts and ends at zero amplitude. This results in approximately 1.87 times larger pulse amplitude for the Gaussian pulse to achieve the same rotation angle as the rectangular pulse with the same gate time. **b**, Spectral comparison of rectangular pulse and Gaussian pulse. The Gaussian pulse exhibits significantly less sidelobe component compared to the rectangular pulse. **c**, Schematic for the implementation of the unconditional IZ readout (Q2 readout) using the PSB parity (ZZ) spin readout. The green arrow indicates the resonance frequency of the PSB reference qubit (Q1) at $-126.6$ MHz. **d**, Randomized benchmarking results for the rectangular and Gaussian pulses with the IZ (red, offset by 0.2) and ZZ (blue) readout. The black solid lines are the fit to an exponential decay $F(m) = Vp^m$, where $V$ is the visibility and $1-p$ is the decay per Clifford gate. **e**, RB decay for various gate time measured with rectangular pulse. The blue band represent the simulation result with $1\sigma$ from the mean, obtained from 100 simulation runs.

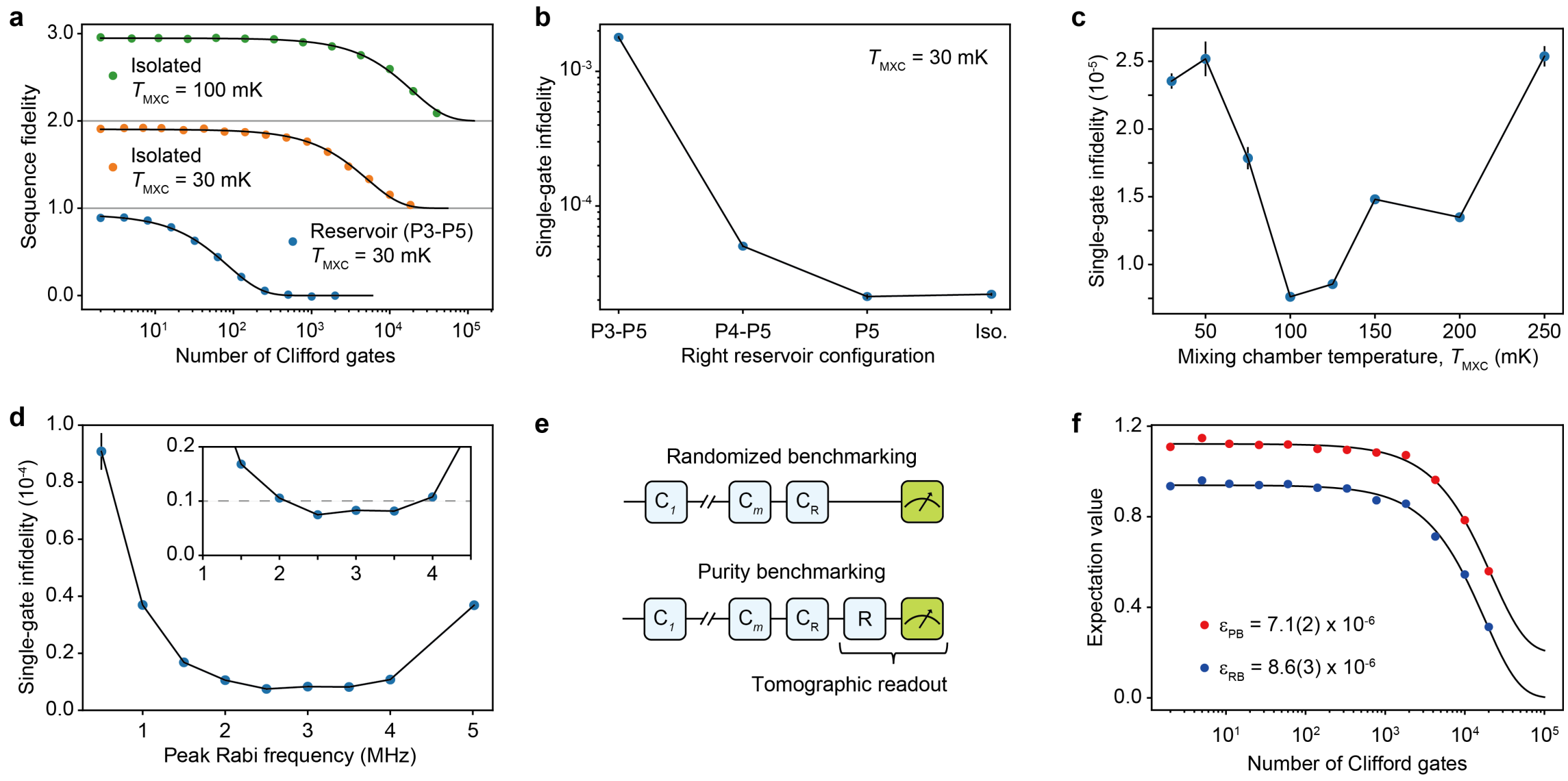

Fig. 5, RB fidelity characterizations. Gaussian pulses with a peak Rabi frequency of 3 MHz are used unless otherwise noted. The error bars represent 1σ from the mean. **a**, Comparison of RB traces measured with different experimental configurations. "Isolated" refers to the configuration where the right reservoir is depleted under gates P3-P5 and B2-B5, while the reservoirs for the charge sensor are still present. The solid black lines are exponential fits with $\pi/2$ gate fidelities of 0.99821(4) (blue), 0.999970(1) (orange), and 0.9999920(2) (green). **b**, RB infidelities measured for different reservoir proximities. The labels for the x-axis refer to the configurations shown in Fig. 2a-c. **c**, RB fidelities measured at different mixing chamber temperatures. **d**, RB infidelities measured at different microwave driving amplitude ($T_{\mathrm{MXC}} = 100$ mK). The inset shows the close-up of the region around the peak Rabi frequency of 3 MHz, where the average power corresponds to a constant $f_{\mathrm{Rabi}}$=1.9 MHz (see Supplementary Information). **e**, Schematic of the RB and PB implementations. In both protocols, a sequence of $m$ random Clifford gates $C_1$, …, $C_m$ is followed by a recovery Clifford gate $C_{\mathrm{R}}$. For RB, a projective readout is performed to obtain the probability projected onto the target state of $C_{\mathrm{R}}$. For PB, a tomographic readout is performed instead: after $C_{\mathrm{R}}$, a pre-rotation $R \in \{\mathrm{I}, \mathrm{X}_{\pi/2}, \mathrm{Y}_{\pi/2}, \mathrm{X}_{\pi}\}$ is applied. The experiment is repeated for all four pre-rotations using the same random Clifford gate sequence, and those measurement outcomes are used to calculate the purity $P$. **f**, Comparison of RB decay (blue) and purity decay (red, offset by 0.2). For the purity, we plot $\sqrt{P}$ instead of $P$ so that it reflects the same amount of depolarizing error as the RB measurement. Solid black lines are exponential fits.

**Supplementary Information for "Assessing fidelity-limiting factors and achieving single-qubit gate fidelity beyond 99.999% in driven silicon spin qubits"**

Capacitive coupling between quantum dot and reservoir

Voltage fluctuations in the reservoir couple to the electrochemical potential of the quantum dot through the electrostatic lever arm between the two structures. Through the micromagnet-induced spin–electric coupling, these voltage fluctuations contribute to the spin-locking relaxation rate $T_{1\rho}^{-1}$. To estimate the distance dependence of this coupling, we model the accumulated 2DEG reservoir as a grounded, infinitesimally thin conducting sheet and approximate the surface gate as a continuous grounded plane, and treat each surface element as a Coulomb kernel to the dot potential. For a surface element of the reservoir at $\boldsymbol{r}$ (with the quantum dot at the origin (0,0,0)), it is given by $G(\boldsymbol{r}) = \frac{1}{4\pi\varepsilon_{\mathrm{eff}}}(|\boldsymbol{r}|^{-1} - |\boldsymbol{r} + (0,0,2t_{\mathrm{eff}})|^{-1})$, where $t_{\mathrm{eff}}$ is the effective dielectric thickness and $\varepsilon_{\mathrm{eff}}$ is the effective dielectric constant; the second term represents the image charge to account for screening by surface gates. To account for charge redistribution within the reservoir, we divide the rectangular 2DEG region into smaller 2 nm x 2 nm square regions. Assuming a total induced charge charge $Q_j$ uniformly distributed over region $j$, the induced charges are obtained by solving the grounded boundary condition at the center $\boldsymbol{r}_i$ of every region:

$$qG(\boldsymbol{r}_i) + \sum_j K_{ij}Q_j = 0, \qquad K_{ij} = \frac{1}{(2\ \mathrm{nm})^2}\int_{A_j} G(\boldsymbol{r}_i - \boldsymbol{r}')dx'dy',$$

where $q$ is the charge at the quantum dot position and $A_j$ is the area of region $j$. The total charge induced in the reservoir defines the dimensionless coupling constant $\eta(d) = -\frac{1}{q}\sum_j Q_j$, and the mutual coupling between the quantum dot and reservoir is given by $C_m(d) = \eta(d)C_{\mathrm{QD}}$, where $C_{\mathrm{QD}}$ is the total capacitance of the quantum dot. For a spatially uniform reservoir-voltage fluctuation, the resulting fluctuation of the quantum dot electrochemical potential is proportional to $\eta(d)^2$. Then, we can obtain the scaling of $T_{1\rho}$ as $T_{1\rho}^{-1} \propto \eta(d)^2 + \mathrm{const.}$, where the constant accounts for the additional distance-independent noise stemming from sources such as the reservoirs for the charge sensor and/or charge noise. Note that the induced charge redistribution gives a negligible correction to the typical 2DEG density of $3 - 5 \times 10^{11}\ \mathrm{cm}^{-2}$.

Supplementary Fig. 9 illustrates the simplified geometry of accumulated 2DEG reservoir and the calculated capacitive coupling as a function of the distance between the quantum dot and the reservoir. Note that while the fit curve is reasonably consistent with the data, significant uncertainty remains due to the small number of points, unknown 2DEG resistivity, spin-electric coupling, and microwave drive-induced heating.

Synchronization condition for rectangular pulse

For an off-resonant qubit separated by a frequency of $\Delta f$, the generalized Rabi frequency is $\sqrt{\Delta f^2 + f_{\text{Rabi}}^2}$, where $f_{\text{Rabi}}$ is the Rabi driving strength of the off-resonant qubit. Therefore, the synchronization condition for a rectangular pulse is

$$2\pi\sqrt{\Delta f^2 + f_{\text{Rabi}}^2} \times t_{\text{g}} = n\pi,$$

where $t_{\text{g}} = 1/(4f_{\text{Rabi}})$ is the $\pi/2$ gate time, and $n$ is a natural number. The off-resonant driving is maximized for odd $n$ and minimized for even $n$. Rewriting the equation for $t_{\text{g}}$ yields the synchronization condition

$$t_{\text{g}}(n) = \frac{\sqrt{4n^2 - 1}}{4|\Delta f|},$$

from which the oscillation period of $t_{\text{g}}(n+2) - t_{\text{g}}(n) \approx |\Delta f|^{-1}$ can be obtained for large $n$. The sampling of $t_{\text{g}}$ used in Fig. 4 is obtained by setting $|\Delta f| = 126.6$ MHz and $n = 12, 13, \ldots, 31$.

Simulation of off-resonant driving effect

The off-resonant time evolution during a single $\pi/2$ gate is calculated using the rotating frame Hamiltonian at the microwave drive frequency (set at the Q2 resonance frequency),

$$H_{\text{x/y}}(t) = \frac{2\pi(\Delta f + \Delta f_{\text{noise}})\sigma_{\text{z}}}{2} + \frac{2\pi\Omega(t)\sigma_{\text{x/y}}}{2}.$$

Here, $\Delta f = -126.6$ MHz is the Q1 resonance frequency offset from the microwave drive frequency, $\Delta f_{\text{noise}}$ is the additional frequency offset caused by qubit-energy fluctuations, and $\Omega(t)$ is the time-dependent Rabi driving amplitude (rectangular or Gaussian) on the off-resonant qubit (Q1). The low-frequency noise term $\Delta f_{\text{noise}}$ is generated by assuming a noise spectrum $S(f) = A^2/f$ and is treated as constant during a single $\pi/2$ gate. The value $A = 17$ kHz is chosen to reproduce the Ramsey $T_2^*$ in Fig. 1d. The time evolution during the waiting time is calculated using $H_w = 2\pi(\Delta f + \Delta f_{\text{noise}}^{\text{w}})\sigma_z/2$, where $\Delta f_{\text{noise}}^{\text{w}}$ is the frequency offset during the waiting time, sampled from the same time trace used for generating $\Delta f_{\text{noise}}$. In the simulation on Fig. 4e, we use a waiting time of 10 ns, identical to the experimental value. From the exponential fit to simulated decay curve, we extract the characteristic decay parameter. Because the simulation does not include the decay originating from the resonant qubit (Q2), we add a constant offset of $2.5 \times 10^{-4}$ (the average for the IZ readout results) to the entire simulated results. The frequency shift due to heating is omitted, as it is significantly smaller than $|\Delta f|$.

IZ readout

The controlled rotation is implemented by first adiabatically turning on the exchange coupling $J$ to ~8 MHz, applying a Gaussian shaped $\pi$ pulse, and then adiabatically turning $J$ off. The exchange pulse is filtered with a Tukey window, accounting for the measured exchange lever arm of 16.9 dec/V for the virtual barrier gate vB1. We do not correct for the phase accumulated during the exchange pulse, as it is irrelevant for the projective measurement.

Amplitude relation between rectangular and Gaussian pulses at equal average power

When comparing heating effects between rectangular pulses and Gaussian pulses, it is relevant to match their average power, which scales with the time-averaged square of the drive amplitude. For a rectangular pulse with an amplitude of $f_{\mathrm{Rabi}}$ and a Gaussian pulse with a peak $f_{\mathrm{Rabi}}$ of $f_{\mathrm{Rabi,peak}}$, this condition corresponds to

$$f_{\mathrm{Rabi}}^2 = \frac{f_{\mathrm{Rabi,peak}}^2}{4\sigma} \int_{-2\sigma}^{2\sigma} \left( \frac{\mathrm{e}^{-\frac{t^2}{2\sigma^2}} - \mathrm{e}^{-2}}{1 - \mathrm{e}^{-2}} \right)^2 dt.$$

Numerically calculating the integral gives the relation $f_{\mathrm{Rabi,peak}} \approx 1.59 f_{\mathrm{Rabi}}$ for matched average power. For example, the rectangular pulse amplitude gives the same power as a Gaussian pulse with 3 MHz peak $f_{\mathrm{Rabi}}$ is approximately 1.9 MHz.

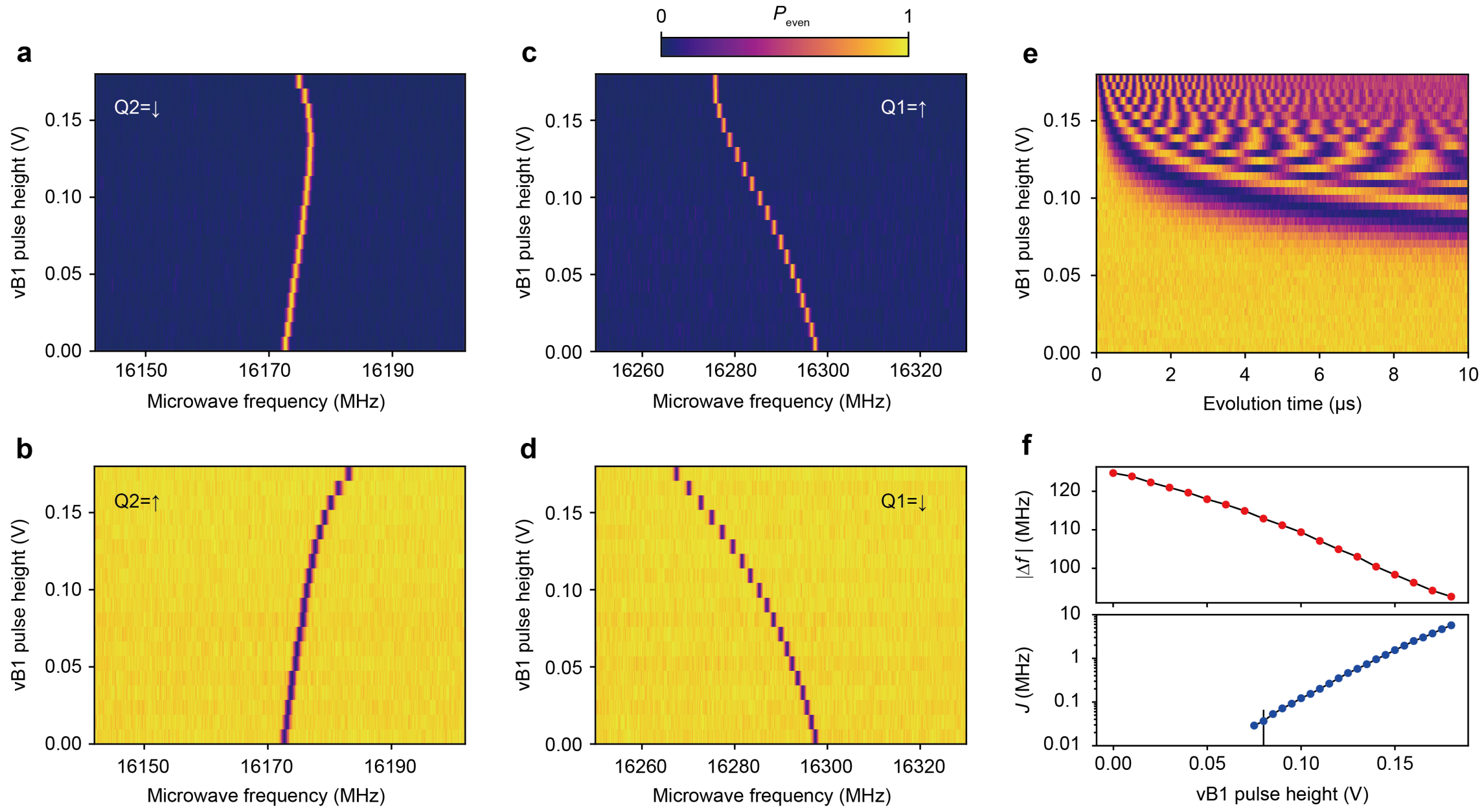

Supplementary Fig. 1, Exchange spectroscopy. **a-d**, Controlled rotations. Panels **a** and **b** show Q1 resonances, while panels **c** and **d** show Q2 resonance peaks. The system is initialized in $|\uparrow\downarrow\rangle$. Then, the spectrum is measured either directly (**a**, **c**) or after applying a $\pi$ flip on the control qubit (**b**, **d**). A low-power Gaussian $\pi$ pulse is used to obtain narrow resonance peaks. **e**, Decoupled controlled phase oscillations measured with Q2 as the target qubit and Q1 as the control qubit. **f,** Qubit frequency difference ($|\Delta f|$) and exchange coupling ($J$) measured as a function of the barrier gate (vB1) pulse height. vB1 pulse height = 0V corresponds to the single-qubit operation configuration. $|\Delta f|$ is extracted from the data in **a-d**, and $J$ is obtained from data in **e**. $|\Delta f|$ decreases approximately linearly as vB1 is increased, whereas $J$ increases exponentially. In the measurement of $J$, no data points are available for pulse heights below 0.075V because no oscillations are observed. Error bars represent $1\sigma$ from the mean.

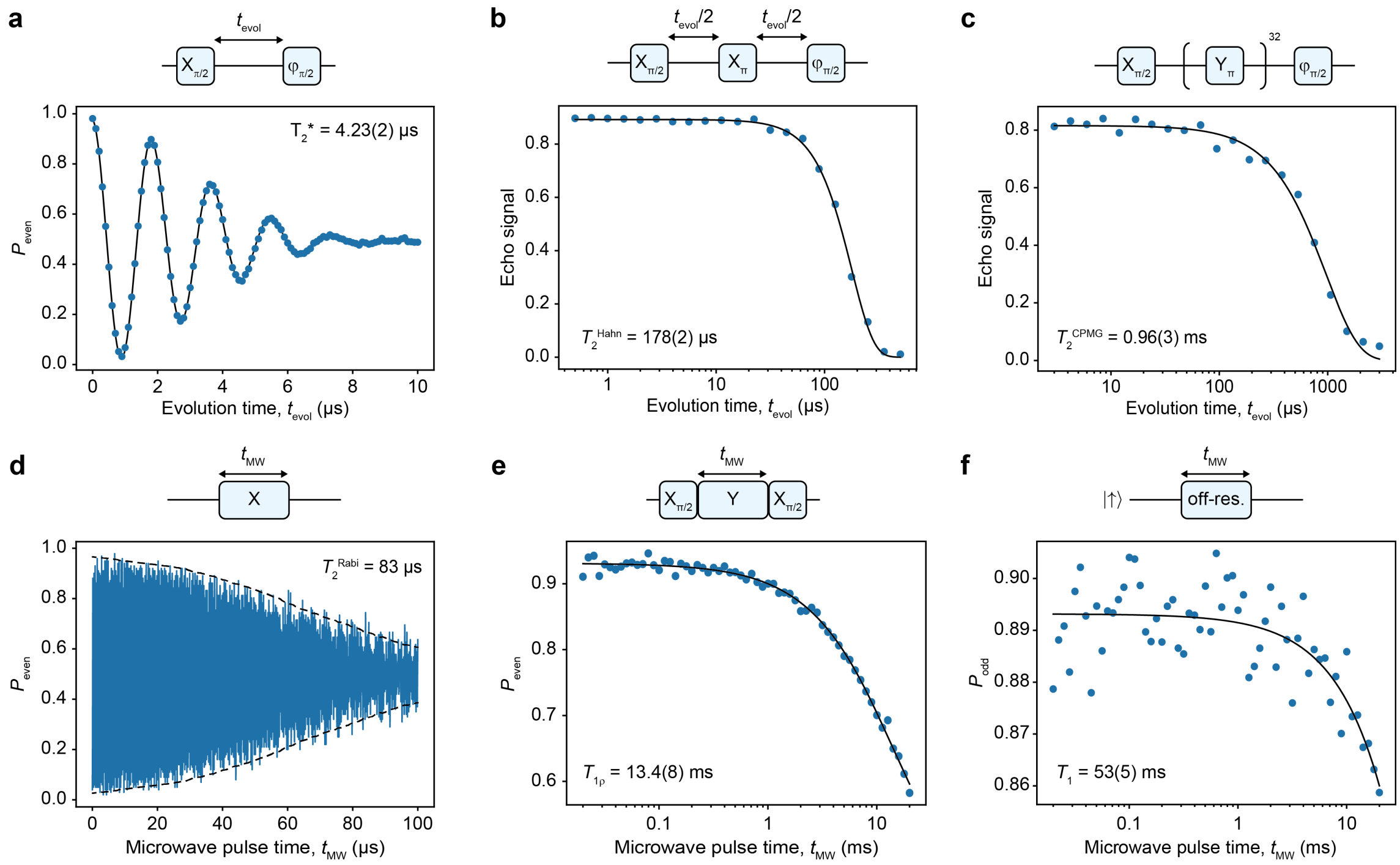

Supplementary Fig. 2, Coherence time measurements for Q2. All pulses have rectangular envelopes, and the Rabi frequency is calibrated to $f_{\mathrm{Rabi}}$ = 3 MHz. All measurements are performed at the base temperature of the dilution refrigerator. **a**, Ramsey $T_2^*$ measurement. A virtual frequency detuning is added to facilitate the curve fitting by varying the phase of the final projection pulse as $\theta = 10\pi/(10\mu s/t_{\mathrm{evol}})$. We average 30,000 shots to obtain a single probability value, resulting in a data acquisition time of approximately 1600 s. Shorter integration times generally yield longer $T_2^*$. The solid line is a fit for Gaussian decay. **b**, Hahn echo measurement. The phase of the final $\pi/2$ pulse is varied from 0 to $2\pi$. The echo signal is then extracted from the oscillation amplitude at a fixed $t_{\mathrm{evol}}$. **c**, CPMG echo measurement. $N = 32$ $Y_{\pi}$ pulses are applied. The evolution time in the plot corresponds to the total evolution time between the pulses. **d**, Rabi decay measurement. The Rabi decay time is defined as the 1/e decay time of the envelope, which is obtained by demodulating the oscillation signal at 3 MHz. **e**, Spin-locking time measurement. The interval between the microwave pulses is 10 ns, which is negligible. The black solid line is a fit to exponential decay. **f**, Relaxation time measurement with an off-resonant microwave pulse. Q2(Q1) is initialized to spin-up (spin-down) prior to the off-resonant pulse application. The off-resonant pulse has an amplitude corresponding to a 3 MHz Rabi frequency. During measurement, the reservoir is extended below gates P3-P5.

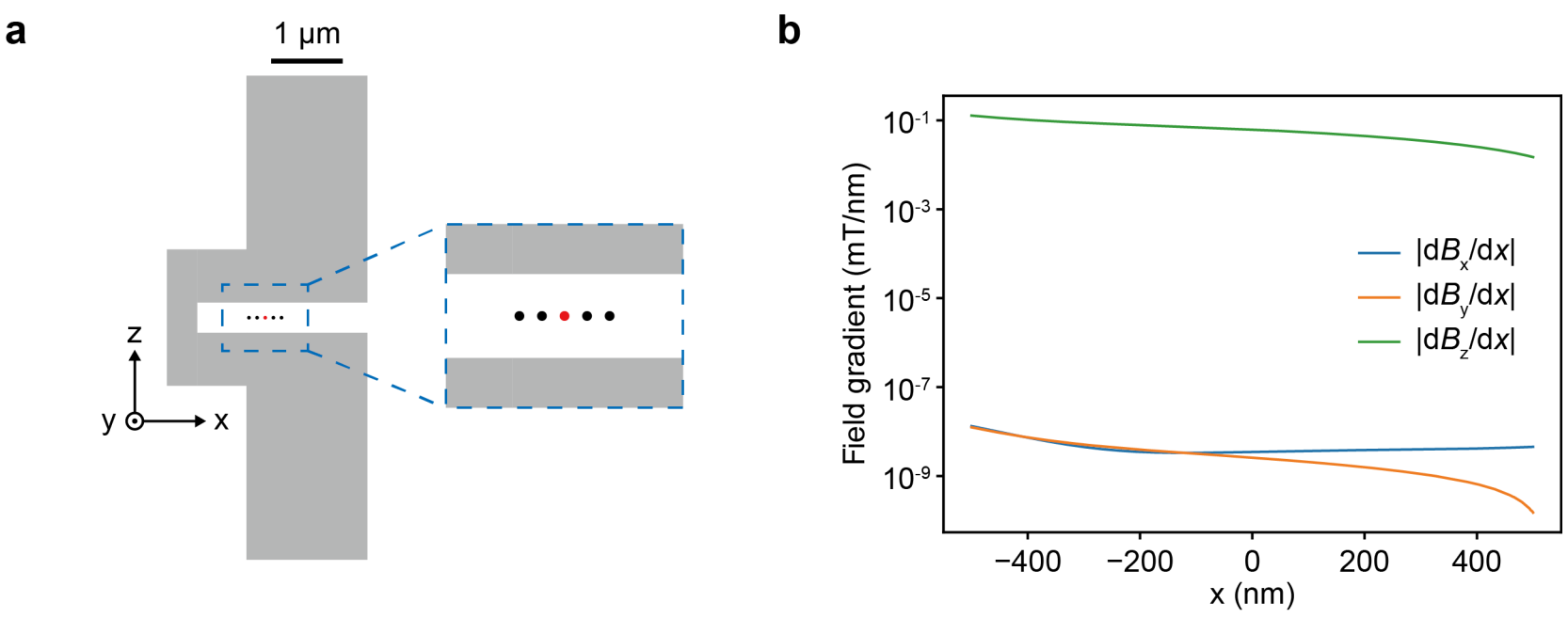

Supplementary Fig. 3, Simulation of micromagnet field gradient anisotropy. **a**, Micromagnet geometry. The red circle, corresponding to QD3, defines the origin of the x- and z-axes. In this design, the QD2 (QD1) position corresponds to $x = -110$ $(-220)$ nm. The distance between the quantum-dot plane and the bottom of the micromagnet is set to 147 nm, which is derived from the average height of the overlapping gate stack. Variations in gate height due to the overlapping gate stack are neglected because they are sufficiently small compared with the average dot-to-magnet distance and are not expected to change the order of magnitude of the resulting field gradients. We assume the cobalt film is fully magnetized with a saturation field of 1.8 T. **b**, Magnetic field gradient along the x-axis at $z = 0$. In the experiment, the quantum dot positions are likely misaligned from $z = 0$ due to disorder; however, such misalignment does not drastically affect the gradients along the x-axis.

**a**

| | Isolated | P3-P5 |
|---|---|---|
| $T_2^*$ (μs) | 3.22(2) | 3.86(2) |
| $T_2^{H}$ (μs) | 90(2) | 186(4) |
| $T_2^{CPMG}$ (μs) | 446(22) | 820(79) |
| $T_2^{Rabi}$ (μs) | 66(11) | 58(11) |
| $T_{1\rho}$ (ms) | 0.85(8) | 0.29(1) |
| $F_p$ | 0.999939(3) | 0.999827(5) |

**b**

$T_{MXC}$=100mK

$\varepsilon_{PB}$ = 3.0(1) x $10^{-5}$

$\varepsilon_{RB}$ = 3.65(9) x $10^{-5}$

Expectation value

Number of Clifford gates

Supplementary Fig. 4, Coherence times and benchmarking for Q1. **a,** Summary of coherence time metrics and primitive gate fidelity measured at 30 mK. $T_{1\rho}$ and $T_2^{\mathrm{Rabi}}$ are measured at $f_{\mathrm{Rabi}}$ = 3 MHz. We use Gaussian pulses for measuring the primitive gate fidelity $F_{\mathrm{p}}$. While the performance for the isolated configuration is lower than that of Q2, it is less sensitive to the reservoir accumulated under P3-P5 due to the larger distance between Q1 and the reservoir. **b**, Comparison of RB and purity benchmarking performed at $T_{\mathrm{MXC}}$ = 100 mK with the isolated configuration.

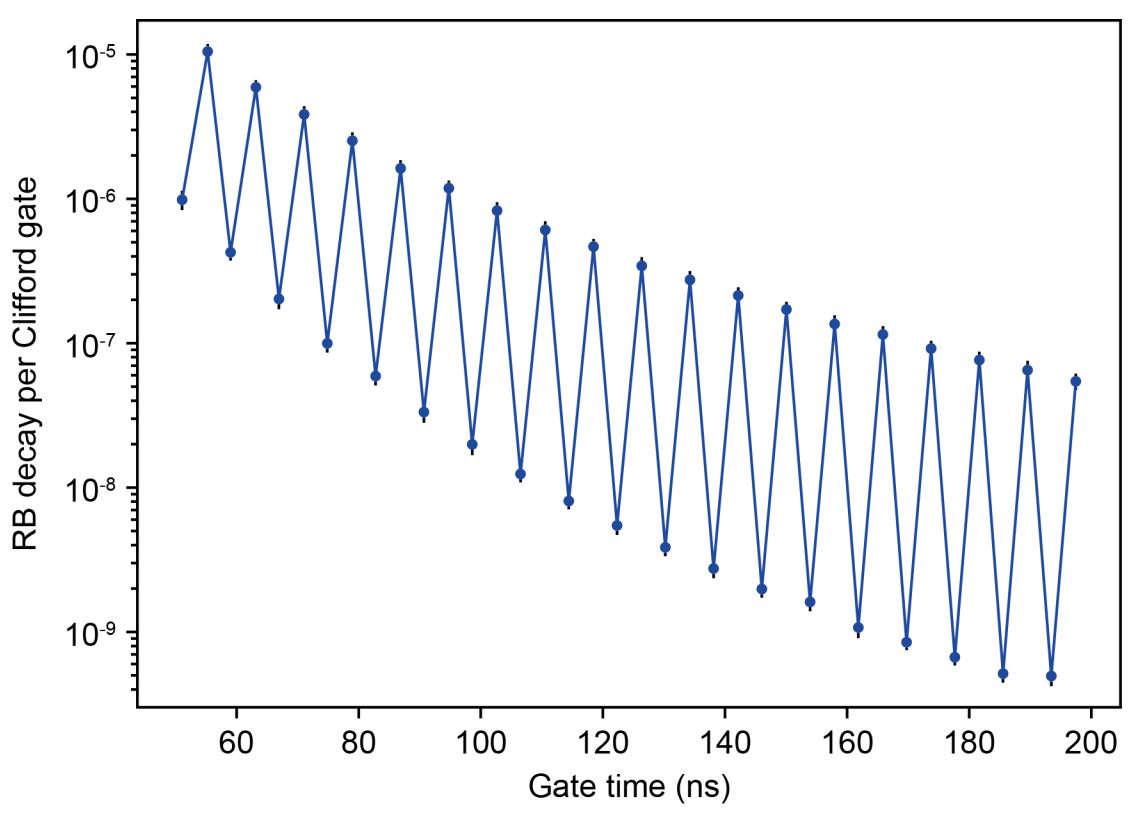

Supplementary Fig. 5, Simulation of RB decay due to off-resonant driving for Gaussian pulses. Pulse truncation results in an oscillation of the decay parameter, which is not expected for infinitely long Gaussian pulses. The peak $f_{\text{Rabi}}$ used in Fig. 5 relates to the gate time as $t_{\text{g}} = 1.87/(4 \times (\text{peak } f_{\text{Rabi}}))$. For $t_{\text{g}}$=83.3 ns (as in Fig. 4d) and 155.8 ns (as in Fig. 5a and 5d), we calculate the decay-per-Clifford-gate values of $8(1) \times 10^{-8}$ and $1.6(2) \times 10^{-8}$, respectively. The errors represent $1\sigma$ from the mean, obtained from 100 simulation runs.

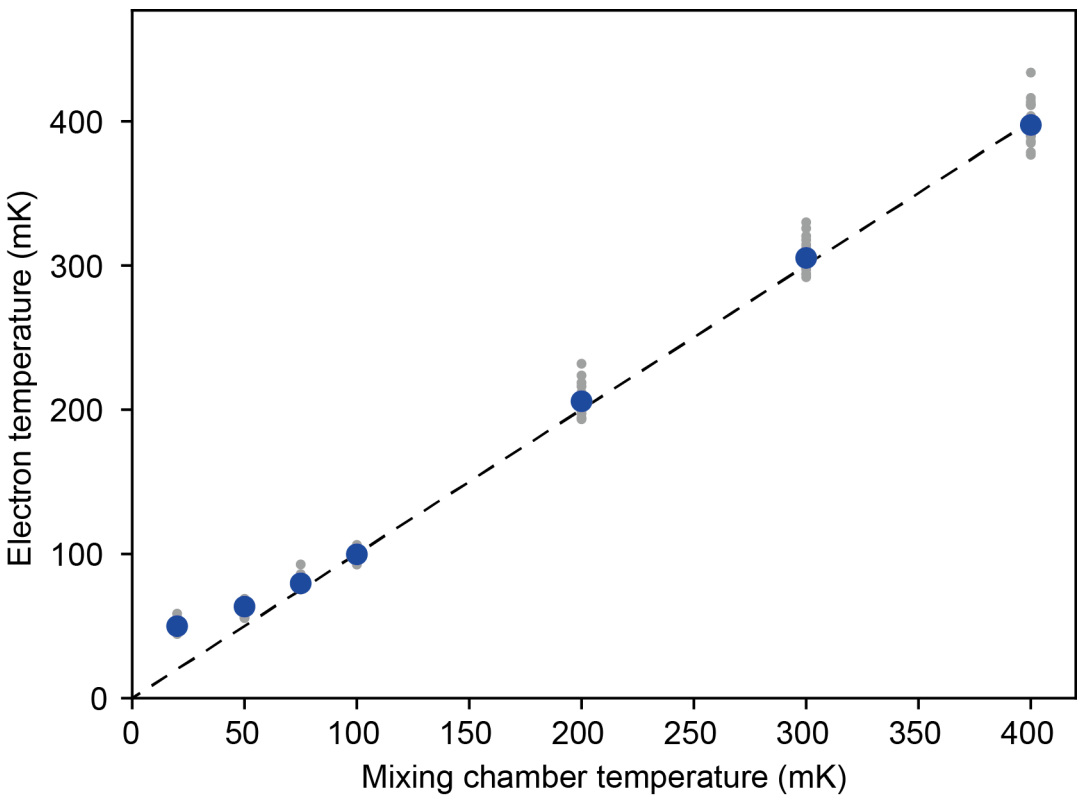

Supplementary Fig. 6, Electron temperature measurement. We monitor the transition width between the quantum dot and the right reservoir (extended to below P3) as a function of the mixing chamber temperature. The external magnetic field is set to 0 T. The blue points are average of 21 measurements, each of which is represented by gray points. The dashed line is an eye guide with a slope of 1. A lever arm of 0.1 eV/V is used to convert the linewidth to the electron temperature.

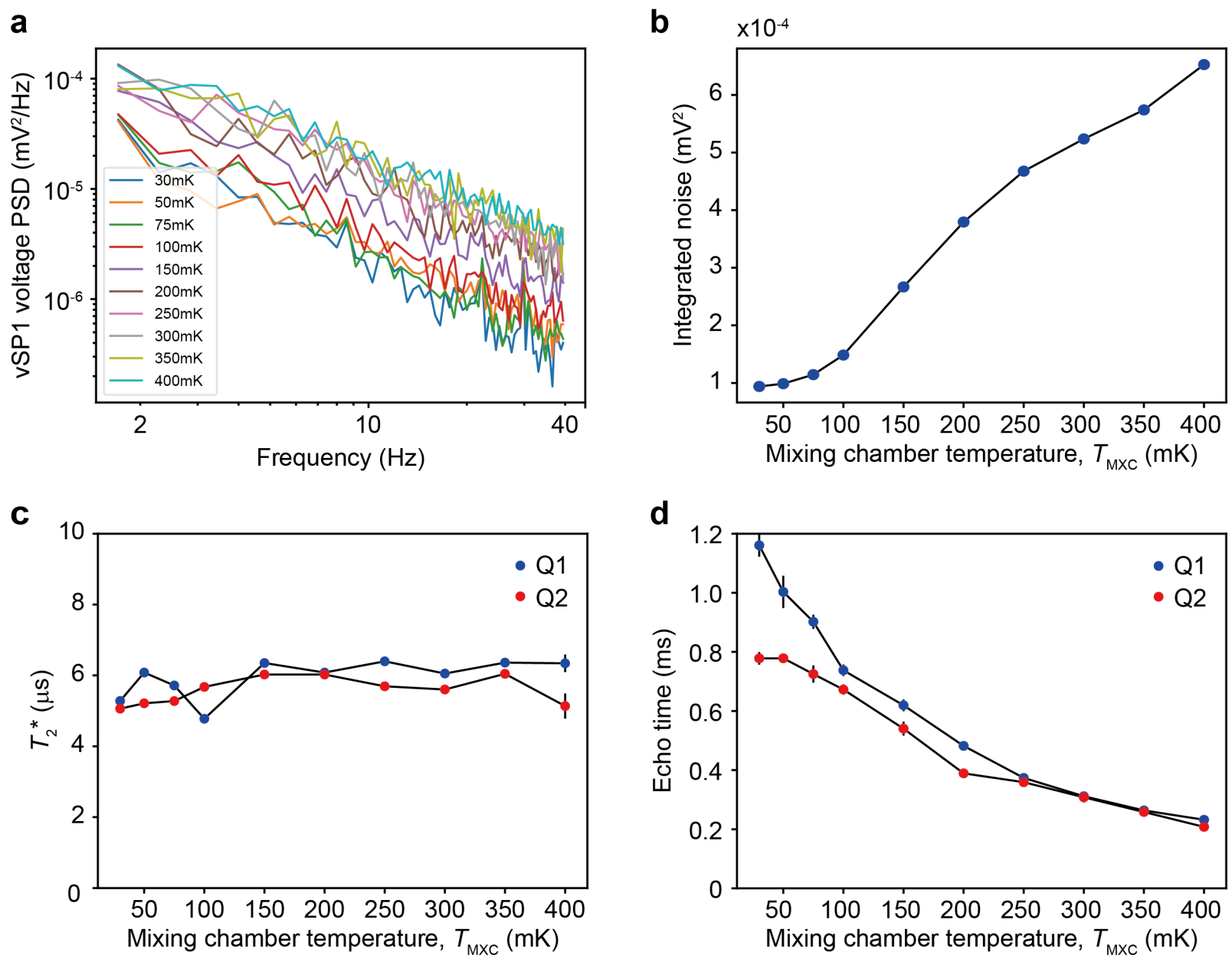

Supplementary Fig. 7, Charge sensor noise and dephasing times measured at various $T_{\mathrm{MXC}}$. **a**, Power spectral density (PSD) of charge sensor signal fluctuations, converted to the SP1 virtual gate voltage. The charge sensor sensitivity, $\mathrm{d}V_{\mathrm{rf}}/\mathrm{d}V_{\mathrm{vSP1}}$, is measured at each temperature and used to convert $V_{\mathrm{rf}}$ into gate voltage fluctuations. The frequency window [1.5 Hz, 40 Hz] is chosen to avoid noise peaks coming from pulse-tube-induced vibrations. **b**, Integrated charge noise, $\int_{1.5}^{40} S(f)df$ as a function of $T_{\mathrm{MXC}}$. The data shows good agreement with the measured $T_{\mathrm{e}}$, with the charge noise power proportional to $T_{\mathrm{e}}$. **c**, Ramsey $T_2^*$ as a function of $T_{\mathrm{MXC}}$. The $T_2^*$ values do not show a significant reduction even if $T_{\mathrm{MXC}}$ is increased, suggesting that $T_2^*$ in this device is not limited by charge noise. The integration time is approximately 108 sec for each data point. **d**, CPMG echo times (32 pulses) as a function of $T_{\mathrm{MXC}}$. Note that these data are measured at a different voltage configuration from that in Fig. 1, resulting in different echo times. The echo dephasing times show a decreasing trend with increasing $T_{\mathrm{MXC}}$.

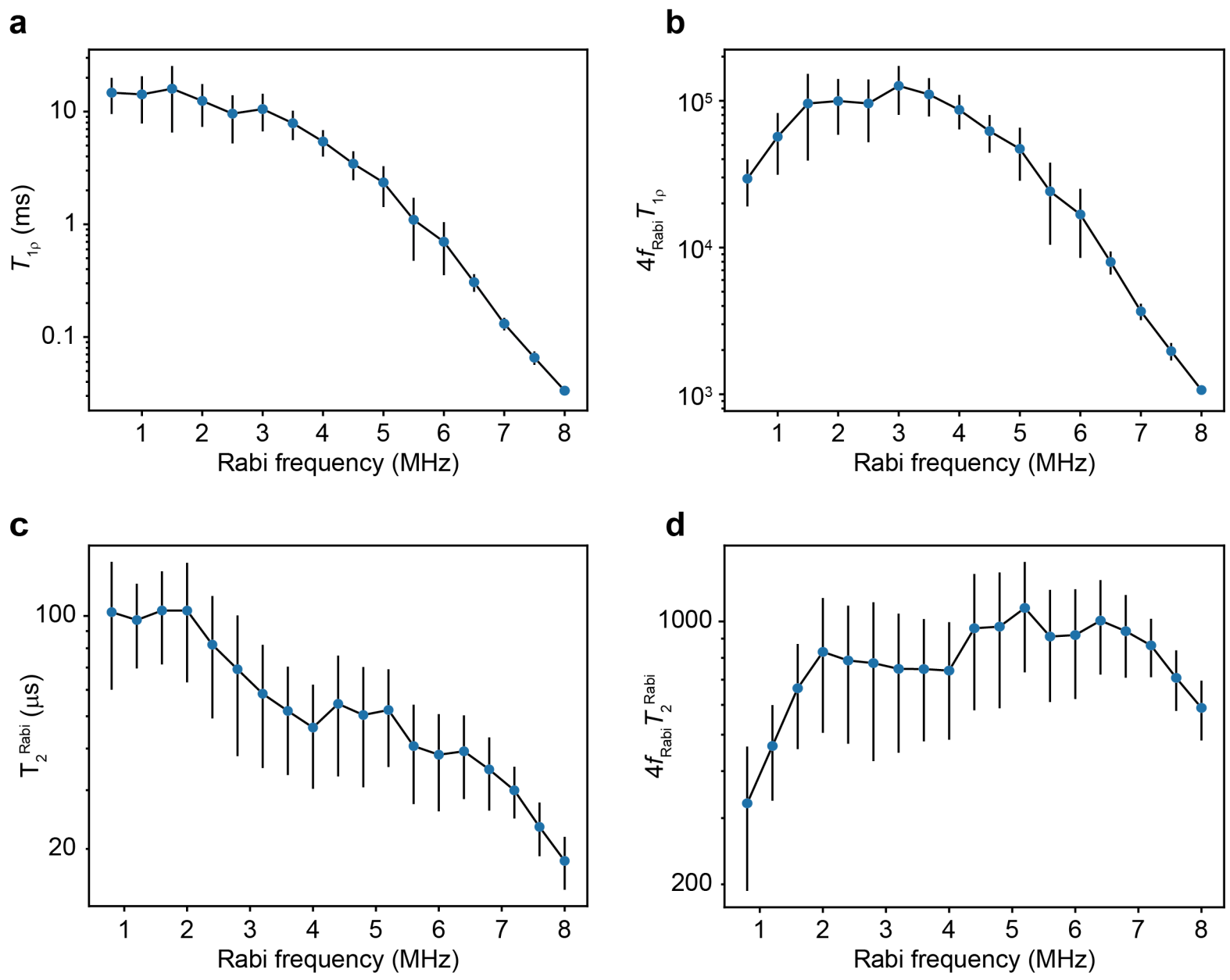

Supplementary Fig. 8, Power dependence of driven coherence times for Q2. All measurements are performed at 30 mK. Error bars represent $1\sigma$ from the mean, extracted from 16 independent measurements. **a**, $T_{1\rho}$ as a function of $f_{\mathrm{Rabi}}$. **b**, $\pi/2$ rotation spin-locking quality factor ($4f_{\mathrm{Rabi}}T_{1\rho}$) as a function of $f_{\mathrm{Rabi}}$. **c,** Rabi decay time as a function of $f_{\mathrm{Rabi}}$. **d**, $\pi/2$ rotation Rabi quality factor ($4f_{\mathrm{Rabi}}T_2^{\mathrm{Rabi}}$) as a function of $f_{\mathrm{Rabi}}$.

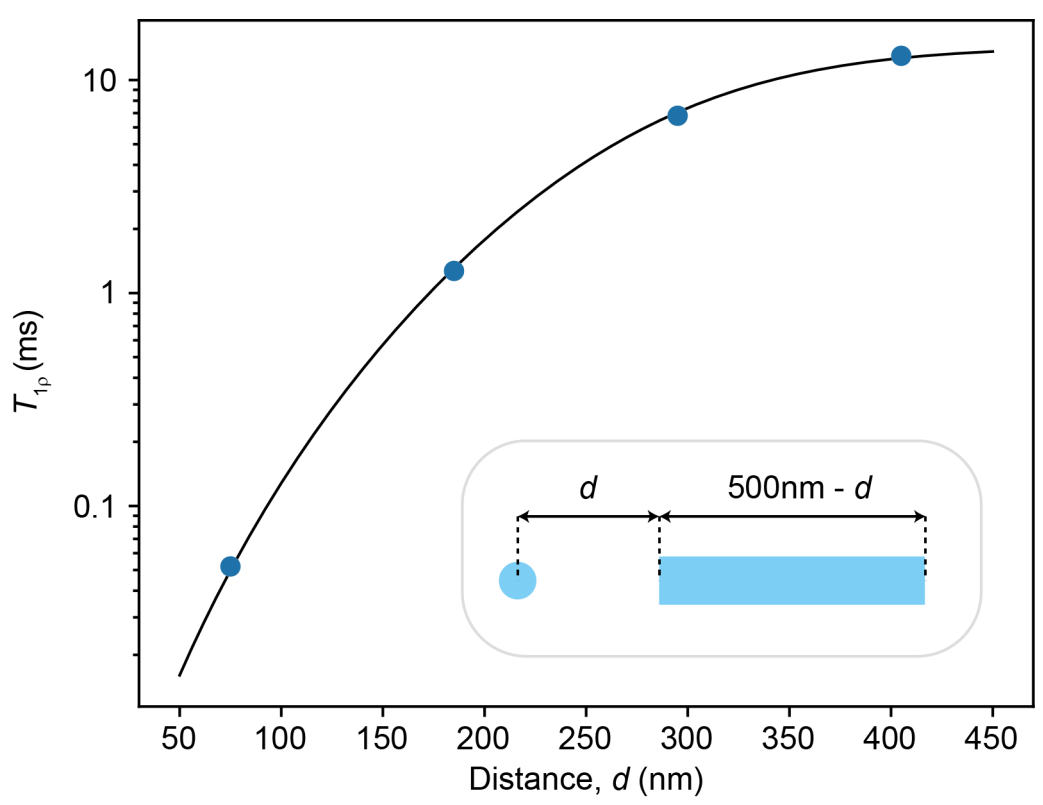

Supplementary Fig. 9, $T_{1\rho}$ and the reservoir distance. The schematic in the inset represents the reservoir geometry used for calculating the capacitance (not to scale; reservoir width is assumed to be 40 nm). The effective dielectric thickness between the gates and the 2DEG ($t_{\mathrm{eff}} = 46.5$ nm, silicon-equivalent) is calculated using $\varepsilon_{\mathrm{Si}} = 11.7$, $\varepsilon_{\mathrm{Al2O3}} = 9$, and $\varepsilon_{\mathrm{Si0.7Ge0.3}} = 13.0$. Circles represent experimental data, and the solid line is the fit curve. The fit curve is calculated as $T_{1\rho}(d)/(1\ \mathrm{ms}) = A/(\eta(d)^2 + B)$, where $A$ and $B$ are dimensionless constants determined from the fit in $\log(T_{1\rho})$ space. Modifying the 500 nm offset in the reservoir geometry changes these constants but does not affect the overall result. The quantum-dot-to-reservoir distance for the experimental data points is defined as the distance between the center of designed quantum dot position and the edge of the gate pattern defining the reservoir. Experimentally, this quantity has a variation due to disorder, which is not taken into account.